\documentclass[preprint,pra,superscriptaddress,amsmath,amssymb]{revtex4-2}
\usepackage{amsmath}
\usepackage{amsfonts}
\usepackage{amssymb}
\usepackage{graphicx}
\usepackage{color}
\usepackage{txfonts}
\usepackage{float}
\usepackage[titletoc]{appendix}
\usepackage[colorlinks={true}]{hyperref}
\hypersetup{citecolor={blue}, filecolor={blue}, linkcolor={blue}, urlcolor={blue}}

\begin{document}
\title{Maximizing orientation of a three-state molecule in a cavity with analytically designed pulses}
\author{Li-Bao Fan}
\affiliation{Hunan Key Laboratory of Nanophotonics and Devices, Hunan Key Laboratory of Super-Microstructure and Ultrafast Process, School of Physics, Central South University, Changsha 410083, China}
\author{Hai-Ji Li}
\affiliation{Hunan Key Laboratory of Nanophotonics and Devices, Hunan Key Laboratory of Super-Microstructure and Ultrafast Process, School of Physics, Central South University, Changsha 410083, China}
\author{Qi Chen}
\affiliation{Hunan Key Laboratory of Nanophotonics and Devices, Hunan Key Laboratory of Super-Microstructure and Ultrafast Process, School of Physics, Central South University, Changsha 410083, China}
\author{Hang Zhou}
\affiliation{Hunan Key Laboratory of Nanophotonics and Devices, Hunan Key Laboratory of Super-Microstructure and Ultrafast Process, School of Physics, Central South University, Changsha 410083, China}
\author{Heng Liu}
\affiliation{Hunan Key Laboratory of Nanophotonics and Devices, Hunan Key Laboratory of Super-Microstructure and Ultrafast Process, School of Physics, Central South University, Changsha 410083, China}
\author{Chuan-Cun Shu}
\email{cc.shu@csu.edu.cn}
\affiliation{Hunan Key Laboratory of Nanophotonics and Devices, Hunan Key Laboratory of Super-Microstructure and Ultrafast Process, School of Physics, Central South University, Changsha 410083, China}

\begin{abstract}
We theoretically explore the precise control of a molecular polariton by strongly coupling the lowest three rotational states of a single molecule with a single-mode cavity. We examine two distinct cavity resonance configurations: a fundamental frequency cavity (\( \omega_c = 2B \) with the rotational constant $B$) resonating with the lowest two rotational states, and a second harmonic cavity (\( \omega_c = 4B \)) coupling with the first and second excited rotational states. We propose two control schemes based on the two polariton configurations and derive the corresponding pulse-area theorems to achieve a theoretical maximum orientation  of 0.7746, identical to the molecule in the absence of the cavity.  The control schemes are analyzed in Carbonyl Sulfide (OCS) molecules in their ground rotational state. Our numerical simulation results demonstrate the theoretical control schemes and analyze the sensitivity of the molecular polariton orientation degree to the control field bandwidth and phases. This work provides a valuable reference for achieving maximum field-free orientation of ultracold three-state molecules in a cavity using analytically designed pulses.
\end{abstract}
\maketitle
\section{Introduction}
Through the strong coupling of molecules and light fields, the internal modes of molecules resonate with the optical cavity modes, forming molecular polaritons with new energy-level structures and properties \cite{hutchison2012modifying,garcia2021manipulating,nagarajan2021chemistry}. This strong coupling effect significantly alters the energy level structure of both molecules and light fields, resulting in new optical properties and quantum dynamic effects \cite{kowalewski2017manipulating,flick2017atoms,tichauer2021multi,hu2020recent,yin2021auger,du2023vibropolaritonic,zou2024amplifying}. These effects include strong coupling phenomena at the single-molecule level \cite{chikkaraddy2016single,ojambati2019quantum}, strong long-range intermolecular interactions mediated by photon-patterned media \cite{xiang2020intermolecular,li2021cavity,nagarajan2021chemistry,xiang2021molecular}, ultrafast Rabi splitting contraction for ultrafast all-optical switches \cite{dunkelberger2018ultrafast,dunkelberger2019saturable}, and enhanced Raman scattering effects observed in vibrational polaritons \cite{shalabney2015enhanced,strashko2016raman,mueller2021surface,esteban2022molecular}.\\ \indent
Ultracold polar molecules, with their rich internal structure of vibration and rotation with long coherence times, and strong coupling with confined cavities, have inspired numerous applications as a promising platform for quantum science and technology \cite{carr2009cold,quemener2012ultracold,cornish2024ultracold,fabri2024impact}. Experimental studies have shown that strongly coupling a molecular vibration to an infrared cavity mode creates molecular vibration polaritons, which significantly alter various physical properties and chemical reactivity in molecular systems  \cite{dunkelberger2016modified,xiang2018two,thomas2019tilting,vergauwe2019modification}. The strong coupling between the rotational states of polar molecules and cavity photons has garnered significant attention. This includes the control of ionization probabilities \cite{fabri2024impact}, the induction of light-induced conical intersections \cite{szidarovszky2018conical,csehi2022competition}, and induce polaritonic relaxation
towards molecular rotations \cite{krupp2024collective}. Due to its significance in various physical and chemical processes \cite{stapelfeldt2003colloquium,ohshima2010coherent,fleischer2012molecular,koch2019quantum,qiang2020echo, Hong2023quantum},  a substantial amount of work has been done in the absence of cavities to create a phenomenon referred to as field-free orientation of molecules using resonant and nonresonant excitation schemes \cite{machholm2001field,daems2005efficient,shu2009carrier,shu2010field,fleischer2011molecular,fleischer2012commensurate,shu2013field,egodapitiya2014terahertz,damari2016rotational,xu2020long,shu2020orientational,hong2021generation,de2009field,znakovskaya2014transition,lin2018all,liu2024unveiling}.
Recently, we proposed a theoretical approach to achieve complete control of molecular polaritons, which involves a single molecule with the lowest two rotational states strongly coupled to a single-mode cavity and driven by pulses \cite{fan2023quantum,fan2023pulse}. We demonstrated that analytically designing a pair of pulses can achieve a maximal orientation of the molecular polariton at 0.577. This naturally leads to a fundamental question: How can we control the molecular polaritons from their initial state to the desired target state to enhance the orientation value? It implies that the model needs to include additional rotational states for the molecules initially in the ground state and derive more complex control schemes to tackle this challenge.\\ \indent
In this work, we extend our previous method to consider the lowest three states of molecules coupled with a single-mode cavity. We explore two distinct cavity resonance coupling configurations. The first one involves a fundamental frequency cavity (\( \omega_c = 2B \) with the rotational constant $B$) resonating with the lowest two rotational states. The second one involves a second harmonic cavity (\( \omega_c = 4B \)) that couples with the first and second excited rotational states. Based on the two polariton configurations, we propose two distinct control schemes and derive the corresponding pulse-area theorems to achieve the theoretical maximum degree of orientation at 0.7746 by calculating the exact amplitudes and phases of the control pulses. Furthermore, we analyze these control schemes in Carbonyl Sulfide (OCS) molecules initially in its ground rotational state. This study offers a valuable reference for maximizing the field-free orientation of an ultracold three-state molecule in a cavity using analytically designed pulses.\\ \indent 
The remainder of this paper is organized as follows: Section \ref{Sec:Model} presents the pulse area theorems for the first and second polariton configurations in detail. In Section \ref{RESULTS}, numerical simulations and discussions are conducted to examine the control schemes for the first and second polariton configurations. Finally, our findings are summarized in Section \ref{conclution}.
\section{Model and Hamiltonian \label{Sec:Model}}
\begin{figure}[htbp]\centering
\resizebox{0.65\textwidth}{!}{%
\includegraphics{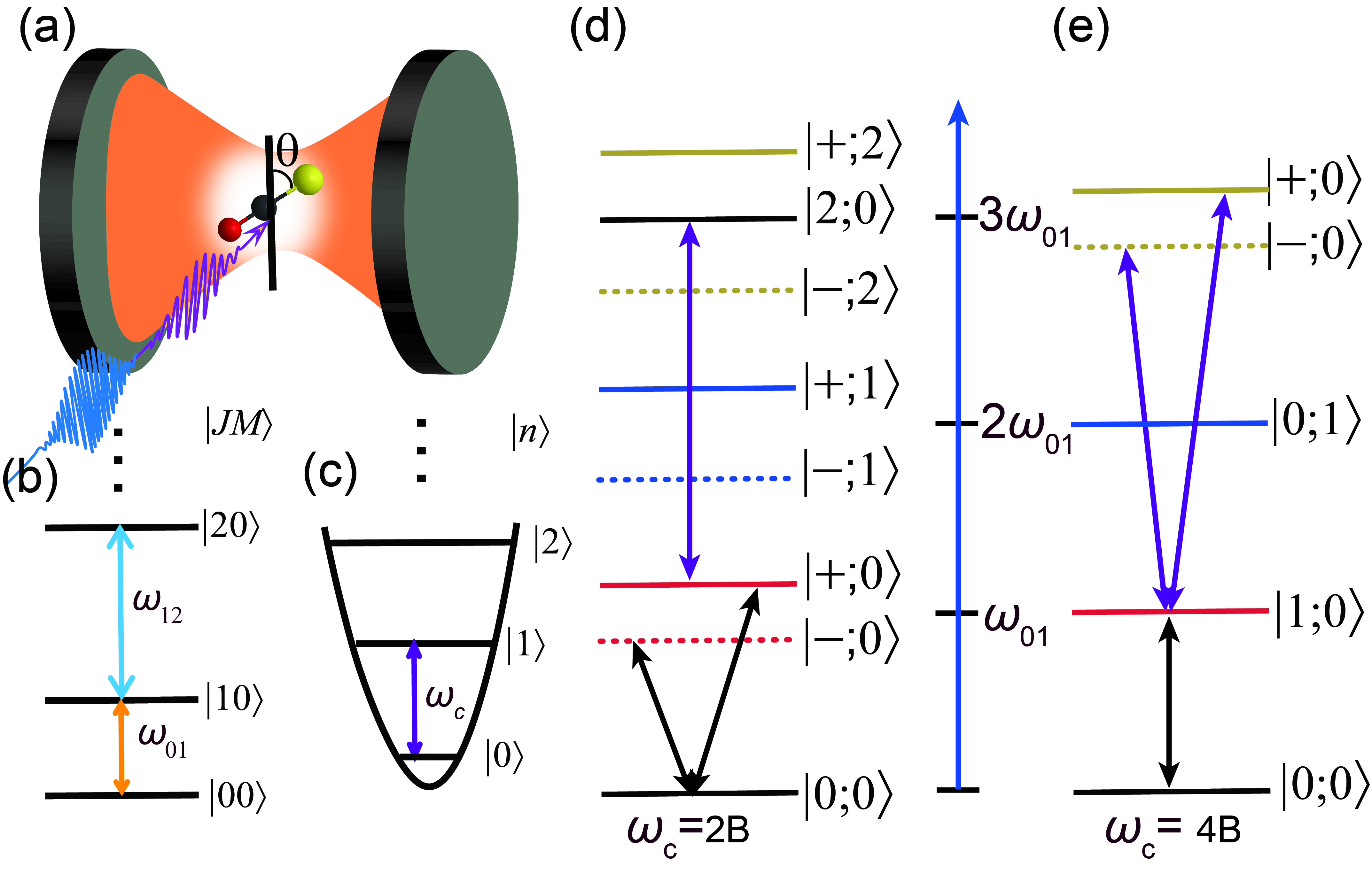}}\caption{(a) Schematic diagram of molecular polaritons driven by composite pulses. (b) Molecular rotation states and (c) cavity field energy level diagram. (d) A fundamental frequency cavity strongly couples with the lowest two states of a rotational three-level single molecule, resulting in hybrid states $|0;0\rangle$, $|\pm;n\rangle$, and $|2;n\rangle$. (e) A second harmonic cavity strongly couples with the first and second excited states of a rotational three-level single molecule, resulting in hybrid states $|0;n\rangle$, $|1;0\rangle$, and $|\pm;n\rangle$.}
\label{fig1}
\end{figure}
We consider a polariton formed by strong coupling of a single molecule with a single-mode cavity, as shown in Fig. \ref{fig1} (a). The  corresponding  Hamiltonian can be written as ($\hbar=1$)
\begin{equation}\label{H}
\Hat{H}_0= B\hat{J}^2+\omega_{c}\hat{a}^{\dagger }\hat{a}-g_0\hat{\mu}\cos\theta\left(\hat{a}+\hat{a}^\dag\right),
\end{equation}
where the first term denotes the field-free Hamiltonian of the molecule with the rotational constant $B$ and the angular momentum operator $\hat{J}$.  The second term corresponds to the cavity Hamiltonian with the creation (annihilation) operator $\hat{a}^{\dagger }$ ($\hat{a}$) and cavity frequency $\omega_{c}$.  The third term describes the interaction Hamiltonian between the molecule and the cavity with $g_{0}=\sqrt{\frac{\omega_c}{2\epsilon_0 V}}$, where $V$ is the volume of the cavity electromagnetic mode, $\epsilon_0$ is the electric constant, the matrix elements of the dipole operator $\hat{\mu}$ read 
$\mu_{JJ’}=\mu\langle J’M\vert\cos\theta\vert JM\rangle$, and the angle $\theta$ between the cavity field polarization and the molecular axis.\\ \indent
%Please explain the model using Figs. 1(b)-(e). What is the model for the molecule and what is a model for the cavity? Additionally, what are the fundamental frequency cavity and the second harmonic cavity? Could you also explain the V2 and 2V excitation/control schemes? Based on these details and the model, you can create a paragraph to guide the readers for reading the next two sections, A and B.
As an extension of our previous work \cite{fan2023pulse,fan2023quantum}, we consider the molecule to consist of the lowest three rotational states and to be initially in its ground state. For the cavity couplings, we examine two different frequency cavities. The first one, referred to as the fundamental cavity with $\omega_c=2B$, resonantly couples with the ground rotational state $J=0$ and the first excited rotational state $J=1$, while the second one, referred to as the second harmonic cavity with $\omega_c=4B$, is resonant with the first and second excited rotational states with $J=1$ and 2. The two different couplings lead to two different polariton configurations, which can be analyzed using an extended quantum Jaynes-Cummings (JC) model \cite{jaynes1963comparison,shore1993jaynes}. The Hamiltonian in Eq. (\ref{H}) for the first polariton configuration  in Fig. \ref{fig1}(d) reduces to  
\begin{equation}
\hat{H}^{JC}_{1}=E_{0,0}\left\vert 0;0\right\rangle \left\langle 0;0\right\vert+\sum_{l=\pm }\sum_{n=0}E_{l,n}\left\vert l;n\right\rangle \left\langle l;n\right\vert+\sum_{n=0}E_{2,n}\left\vert 2;n\right\rangle \left\langle 2;n\right\vert
\end{equation}
where  $\vert0;0\rangle=\vert00\rangle\vert0\rangle$ denotes the vacuum ground state  with the eigenenergy $E_{0,0}=0$, $\vert\pm;n\rangle=\sqrt{2}/2\left(\vert00\rangle\vert n+1\rangle\pm\vert10\rangle\vert n\rangle\right)$ correspond to the  maximally entangled polariton states with the  eigenenergies $E_{\pm,n}=2B(n+1)\pm g_{1}\sqrt{n+1}$, and $\vert 2;n\rangle=\vert 20\rangle \vert
n\rangle$ are the direct-product states with the eigenenergies $E _{2,n}=(2n+6)B$. The coupling strength between the $\vert00\rangle$ and $\vert10\rangle$ states can be calculated by $g_{1}=g_0\mu_{01}$ with $\mu_{01}=\langle 00\vert\mu\cos\theta\vert10\rangle=\sqrt{1/3}\mu$.\\ \indent
The Hamiltonian in Eq. (\ref{H}) for the second polariton configuration in Fig. \ref{fig1}(e) reads
\begin{equation}
\hat{H}^{JC}_2=E_{0,0}\left\vert 0;0\right\rangle \left\langle 0;0\right\vert+E_{1,0}\left\vert
1;0\right\rangle \left\langle 1;0\right\vert+\sum_{l=\pm }\sum_{n=0}E_{l,n}\left\vert l;n\right\rangle\left\langle l;n\right\vert +\sum_{n=1}E_{0,n}\left\vert
0;n\right\rangle \left\langle 0;n\right\vert ,
\end{equation}
where $\vert 1;0\rangle=\vert 10\rangle \vert 0\rangle$ represent the first excited state with the eigenenergy $E_{1,0}=2B$,  $\vert\pm;n\rangle=\sqrt{2}/2(\vert10\rangle\vert n+1\rangle\pm\vert20\rangle\vert n\rangle)$ correspond to the maximally entangled polariton states with the eigenenergies $E_{\pm,n}=(2n+6)B\pm g_{2}\sqrt{n+1}$, and $\vert0;n\rangle=\vert00\rangle\vert n\rangle$  are the direct-product states
with the eigenenergies $E_{0,n}=4nB$. The coupling strength between the $\vert10\rangle$ and states $\vert20\rangle$ can be calculated by $g_{2}=g_{0}\mu_{12}$ with $\mu_{12}=\langle 10\vert\mu\cos\theta\vert20\rangle=\sqrt{4/15}\mu$.\\ \indent 
 Since the orientation values of molecular polaritons depend only on the rotational states $J$ of the molecule \cite{fan2023pulse,fan2023quantum}, the maximum degree of orientation can be achieved with a minimal model that contains all the rotational states $J$ of the molecule and involves the least number of photons. Thus, a minimal four-level system consisting of $\vert 0;0\rangle$, $\vert \pm;0\rangle$, and $\vert 2;0\rangle$ is used for the first polariton configuration. To control a desired superposition state within the four-level system, we employ a $V2$ control scheme, which involves a $V$-type three-level excitation from the ground vacuum state $\vert 0;0\rangle$ to the states $\vert\pm;0\rangle$ following by a two-level excitation from the state $\vert +;0\rangle$ to $\vert 2;0\rangle$. Similarly, the second polariton configuration can be simplified to a four-level system consisting of $\vert 0; 0 \rangle$, $\vert 1; 0 \rangle$, and $\vert \pm; 0 \rangle$, for which we use a 2$V$ control scheme by involving a two-level excitation from $|0;0\rangle$ to $|1;0\rangle$ and a $V$-type three-level excitation from  $\vert 1;0\rangle$ to $\vert \pm;0\rangle$. In the following, we will derive the corresponding pulse-area theorems to design the optimal control pulses for achieving the maximum orientation values for the two polariton configurations.
\subsection{Pulse-area theorem for the first polariton configuration \label{V2}}
For the first polariton configuration,  we define  the desired coherent superposition states of the four-level system  at the target time $t_f$ as
\begin{equation} \label{WFp}
\vert\psi_1(t_{f})\rangle =C_{0,0}(t_{f})\vert
0;0\rangle +\sum_{l=\pm }C_{l,0}(t_{f})e^{-i\omega
_{l,0}t_{f}}\vert l;0\rangle +C_{2,0}(t_{f})e^{-i\omega_{2,0}t_{f}}\vert 2;0\rangle,
\end{equation}
where $C_{0,0}(t_{f})$, $C_{l,0}(t_{f})$, and $C_{2,0}(t_{f})$ represents the complex coefficient of the groud state $\vert 0;0\rangle$, maximally entangled polariton states $\vert\ell;0\rangle$, and direct-product state $\vert 2;0\rangle$, respectively. The parameters $\omega_{\pm,0}=E_{\pm,0}$ and $\omega_{2,0}=E_{2,0}$ represent the eigenfrequencies of the states $\vert \pm;0\rangle$ and $\vert 2;0\rangle$, respectively.\\ \indent
The orientation degree of the four-state system after rotational excitation can be expressed as
\begin{eqnarray}\label{cs1}
\langle\cos\theta(t_{f})\rangle&=&2\sum_{\ell=\pm }\vert C_{0,0}(t_{f})\vert \vert C_{\ell,0}(t_{f})\vert
\cos (-\omega _{\ell,0}t_{f}+\phi _{\ell,0})M_{\ell,0} \nonumber\\
&&+2\sum_{\ell=\pm}\vert C_{\ell,0}(t_{f})\vert \vert C_{2,0}(t_{f})\vert \cos
\left[(\omega _{\ell,0}-\omega_{2,0})t_{f}+\phi _{2,\ell}\right]M_{2,\ell},
\end{eqnarray}
where the phases $\phi _{\ell,0}=\arg [C_{\ell,0}(t_{f})]-\arg [C_{0,0}(t_{f})]$ and $\phi _{2,\ell}=\arg [C_{2,0}(t_{f})]-\arg [C_{\ell,0}(t_{f})]$. The transition matrix $M_{\ell,0}=\langle \ell;0\vert\cos\theta\vert 0;0\rangle$ and $M_{2,\ell}=\langle 2;0\vert\cos\theta\vert \ell;0\rangle$.\\ \indent
Based on the Lagrange multiplier method \cite{Sugny2004Reaching,Wang2020Optimal}, the maximum degree of orientation can
be obtained
\begin{eqnarray}\label{MDO1}
f(\vert C_{0,0}(t_{f})\vert,\vert C_{-,0}(t_{f})\vert,\vert C_{+,0}(t_{f})\vert,\vert C_{2,0}(t_{f})\vert)=\lambda_{1} &=& \sqrt{M_{-,0}^{2}+ M_{+,0}^{2}+M_{2,-}^{2}+ M_{2,+}^{2}}=\sqrt{\frac{3}{5}},
\end{eqnarray}
where the amplitude satisfies $\vert C_{0,0}(t_{f})\vert=\sqrt{10}/6$, $\vert C_{-,0}(t_{f})\vert=\vert C_{+,0}(t_{f})\vert=1/2$, $\vert C_{2,0}(t_{f})\vert=\sqrt{2}/3$, and the phase conditions satisfy the following relationship
\begin{eqnarray}\label{MDOl1}
-\omega_{+,0}t_{f}+\phi_{+,0}&=&2k\pi,\nonumber\\
-\omega_{-,0}t_{f}+\phi_{-,0}&=&\pi+2k\pi,\nonumber\\
(\omega_{+,0}-\omega_{2,0})t_{f}+\phi_{2,+}&=&2k\pi,\nonumber\\
(\omega_{-,0}-\omega_{2,0})t_{f}+\phi_{2,-}&=&\pi+2k\pi.
\end{eqnarray}\\ \indent
To obtain such desired target state based on the $V2$ control scheme, we apply two pulses $\mathcal{E}_{\pm}(t)$ and a time-delayed pulse $\mathcal{E}_{2,+}(t)$ to controlling the system from the ground state $\vert 0;0\rangle$ to the states $\vert +;0\rangle$ and $\vert -;0\rangle$, and the direct-product state $\vert 2;0\rangle$. The corresponding time-dependent Hamiltonian can be given by  
\begin{eqnarray}\label{hp1}
\hat{H}_{p1}(t) &=&\hat{H}^{JC}_{1}- \sum_{\ell=\pm }\mathcal{E}_{\ell}(t)\tilde{\mu}_{0}(
\vert \ell;0\rangle\langle 0;0\vert+\vert 0;0\rangle\langle \ell;0\vert) \nonumber \\
&&-\mathcal{E}_{2,+}(t)\mu_{2,+}(\vert +;0\rangle\langle 2;0\vert +\vert 2;0\rangle \langle+;0\vert),
\end{eqnarray}
where $\tilde{\mu}_0=(\pm\sqrt{2}/2)\mu_{01}$ and $\mu_{2,+}=(\sqrt{2}/2)\mu_{12}$ being the transition dipole moments between the four- state.\\ \indent
%, with $\mu_{01}=\langle 00|\mu\cos\theta\vert10\rangle=\sqrt{3}/3\mu$ and $\mu_{12}=\langle 10\vert\mu\cos\theta\vert20\rangle=\sqrt{4/15}\mu$.
 To obtain the time-dependent wave function by the $V2$ excitation, we use the unitary evolution operator $\hat{U}(t,t_{0})=\mathcal{I}-i\int_{t_{0}}^{t}d t^{\prime}\hat{H}_{I}(t^{\prime})\hat{U}(t^{\prime},t_{0})$ with $\hat{H}_{I}(t)=\text{exp}(i\hat{H}_{1}^{JC}t)(\hat{H}_{p1}(t)-\hat{H}_{1}^{JC})\text{exp}(-i\hat{H}_{1}^{JC}t)$. By applying the first-order Magnus expansion to the unitary evolution operator \cite{blanes2009magnus},  the analytical wave function of the four-level system via $V2$ excitation  can be expressed as
\begin{eqnarray}
\label{Vtype}
\vert\psi_1^{(1)}(t)\rangle &=&\cos\theta_{0}(t)\vert0;0\rangle+\frac{i\theta_{-,0}^{*}(t)}{
\theta_{0}(t)}\sin\theta_{0}(t)\vert-;0\rangle\nonumber\\
&&+\frac{i\theta_{+,0}^{*}(t)}{\theta
_{0}(t)}\sin\theta_{0}(t)\left[\cos \theta _{1}(t)\vert +;0\rangle
+\frac{i\theta _{1}^{*}(t)}{\vert\theta _{1}(t)\vert}\sin \theta _{1}(t)\left\vert
2;0\right\rangle\right],
\end{eqnarray}
with the complex pulse areas $\theta_{\pm,0}(t)=\tilde{\mu}_0\int_{t_{0}}^{t}dt^{\prime }\mathcal{E}_{\pm}(t^{\prime })e^{i(\omega_{\pm,0})t^{\prime}}$ and $\theta_{0}(t)=\sqrt{\vert\theta_{+,0}(t)\vert^2+
\vert\theta_{-,0}(t)\vert^2}$, and $\theta_{1}(t)=\mu_{2,+}\int_{t_{0}}^{t}dt^{\prime }\mathcal{E}_{2,+}(t^{\prime })e^{i(\omega_{2,0}-\omega_{+,0})t^{\prime}}$.\\ \indent 
By comparing Eq. (\ref{Vtype}) with the desired wave function of the target states in Eq. (\ref{WFp}), the maximum degree of orientation in Eq. (\ref{cs1}) can be obtained with optimal amplitude conditions 
\begin{eqnarray}\label{am1}
\vert\theta_{+,0}(t_{f})\vert&=&\sqrt{\frac{17}{26}}\arccos\frac{\sqrt{10}}{6},\nonumber\\
\vert\theta_{-,0}(t_{f})\vert&=&\sqrt{\frac{9}{26}}\arccos\frac{\sqrt{10}}{6},\nonumber\\
\vert\theta_{1}(t_{f})\vert&=&\arccos\frac{3}{\sqrt{17}},
\end{eqnarray}
and the relative phase conditions
\begin{eqnarray}\label{sp1}
-\omega_{+,0}\arg[\theta_{-,0}(t_{f})]+\omega_{-,0}\arg[\theta_{+,0}(t_{f})]=2g_{1}k\pi,\nonumber\\
-(\omega_{2,0}-\omega_{+,0})\arg[\theta_{+,0}(t_{f})]+\omega_{+,0}\arg[\theta_{1}(t_{f})]=2g_{1}k\pi.
\end{eqnarray}\\ \indent
By performing the frequency-domain analysis of three control pulses, we can obtain the relations of $\theta_{\pm,0}^{*}(t_{f})=\tilde{\mu}_{0}A_{\pm}(\omega_{\pm,0})e^{i\phi_{\pm}(\omega_{\pm,0})}$ and $\theta_1^{*}({t_f})=\mu_{2,+}A_{2,+}(\omega_{2,+})e^{i\phi_{2,+}(\omega_{2,+})}$ with the spectral amplitudes $A_{\pm}(\omega_{\pm,0})$ and $A_{2,+}(\omega_{2,+})$, and phases $\phi_{\pm}(\omega_{\pm,0})$ and $\phi_{2,+}(\omega_{2,+})$ at the transition frequencies $\omega_{\pm,0}$ and $\omega_{2,+}=\omega_{2,0}-\omega_{+,0}$. As an example, we consider the spectral amplitudes of three control pulses in Gaussian-frequency distributions $A_{\pm}(\omega_{\pm,0})=A_{\pm,0}\text{exp}[{-(\omega-\omega_{\pm})^2/2(\Delta\omega)^2}]$ and $A_{2,+}(\omega_{2,+})=A_{2,+}\text{exp}[{-(\omega-\omega_{2,+})^2/2(\Delta\omega)^2}]$ with $A_{\pm,0}=\vert\theta_{\pm,0}(t_{f})\vert/\tilde{\mu}_{0}$ and $A_{2,+}=\vert\theta_{1}(t_{f})\vert/\mu_{2,+}$, and set the values of spectral phases $\phi_{\pm}(\omega_{\pm,0})=\varphi_{\pm,0}-\omega\tau_{\pm}$ and $\phi_{2,+}(\omega_{2,+})=\varphi_{2,+}-\omega\tau_{2,+}$.  By  performing  inverse Fourier transforms, we can obtain the three time-dependent control pulses as 
\begin{eqnarray} \label{w1e1}
\mathcal{E}_{\pm}(t)&=&\mathcal{E}_{\pm}
e^{-\frac{(t-\tau_{\pm})^2}{2\tau^2}} \cos[\omega_{\pm,0}(t-\tau_{\pm}) + \varphi_{\pm,0}],  \nonumber \\
\mathcal{E}_{2,+}(t)&=& \mathcal{E}_{2,+} e^{-\frac{(t-\tau_{2,+})^2}{2\tau^2}} \cos[(\omega_{2,0}-\omega_{+,0})(t-\tau_{2,+}) + \varphi_{2,+}],
\end{eqnarray}
with the strengths of the electric fields  $\mathcal{E}_{\pm}=\sqrt{\frac{2}{\pi}} \frac{1}{\tau}   \frac{|\theta_{\pm,0}|}{\tilde{\mu}_{0}}$
and $\mathcal{E}_{2,+}=\sqrt{\frac{2}{\pi}} \frac{1}{\tau}\frac{\vert\theta_{1}\vert}{\mu_{2,+}}$, the pulse duration $\tau=1/\Delta\omega$. By setting $\varphi_{\pm,0}-\omega_{\pm,0}\tau_{\pm}=\phi_{l,0}-\frac{\pi}{2}$ and $\varphi_{2,+}-(\omega_{2,0}-\omega_{+,0})\tau_{2,+}=\phi_{2,+}-\frac{\pi}{2}$, we can verify that the designed control pulses in Eq. (\ref{w1e1}) satisfies both the amplitude and phase conditions in Eqs. (\ref{am1}) and (\ref{sp1}). The details of how we use time-frequency analysis to design optimal control fields can be found in our previous works \cite{fan2023pulse,fan2023quantum,shu2020orientational,hong2021generation,  guo2022cyclic,gong2022discrimination,guo2024optimal}.
\subsection{Pulse-area theorem for the second polariton configuration}
For the second polariton configuration, the wave function $\vert\psi_{2}(t_{f})\rangle$ of the four-state can be written as
\begin{equation} \label{WFp2}
\vert\psi_2(t_{f})\rangle=C_{0,0}(t_{f})\vert 0;0\rangle+C_{1,0}(t_{f})e^{-i\omega_{1,0}t_{f}}\vert 1;0\rangle+\sum_{l=\pm
}C_{l,0}(t_{f})e^{-i\omega_{l,0}t_{f}}\vert l;0\rangle,
\end{equation}
where $C_{0,0}(t_{f})$ denote the complex coefficients of states $\vert0;0\rangle$, $C_{1,0}(t_{f})$ denotes the complex coefficients of the state $\vert 1;0\rangle$, and $C_{\pm,0}(t_{f})$ correspond to the complex coefficients of the doublet states $\vert\pm;0\rangle$. The eigenfrequencies $\omega_{1,0}=2B$ and $\omega_{\pm,0}=6B\pm g_{2}$.\\ \indent
The degree of orientation for the molecular polariton after the rotational excitation can be obtained as
\begin{eqnarray}\label{cs2}
\langle\cos\theta (t_{f})\rangle_{2}&=&2\vert C_{0,0}(t_{f})\vert \vert C_{1,0}(t_{f})\vert\cos(
-\omega_{1,0}t_{f}+\phi_{1,0}) M_{1,0} \nonumber\\
&&+2\sum_{l=\pm }\vert C_{1,0}(t_{f})\vert \vert C_{l,0}(t_{f})\vert
\cos[(\omega _{1,0}-\omega _{l,0}) t_{f}+\phi _{1,l}]M_{1,l},
\end{eqnarray}
where $M_{1,0}=\langle 1;0\vert\cos\theta\vert 0;0\rangle$ and $M_{1,l}=\langle 1;0\vert\cos\theta\vert l;0\rangle$, and the relative phases are $\phi_{1,0}=\arg [C_{1,0}(t_{f})]-\arg [C_{0,0}(t_{f})]$, and $\phi_{1,l}=\arg [C_{l,0}(t_{f})]-\arg [C_{1,0}(t_{f})]$.\\ \indent
The maximum  degree of orientation can be obtained as
\begin{eqnarray}\label{MDO2}
\lambda_{2} &=& \sqrt{M_{1,0}^{2}+ M_{1,+}^{2}+M_{1,-}^{2}}=\sqrt{\frac{3}{5}},
\end{eqnarray}
where $\vert C_{0,0}(t_{f})\vert = \sqrt{10}/6, \vert C_{1,0}(t_{f})\vert=\sqrt{2}/2, \vert C_{\pm,0}(t_{f})\vert=1/3$, the phases satisfy
\begin{eqnarray}\label{ph2}
-\omega_{1,0}t_{f}+\phi_{1,0}&=&2k\pi,\nonumber\\
(\omega_{1,0}-\omega_{-,0})t_{f}+\phi_{1,-}&=&\pi+2k\pi,\nonumber\\
(\omega_{1,0}-\omega_{+,0})t_{f}+\phi_{1,+}&=&2k\pi. 
\end{eqnarray}
 The corresponding Hamiltonian for  the second polariton configuration via the $2V$ excitation can be given by 
\begin{eqnarray} \label{hp2}
\hat{H}_{p2}(t) 
&=&\hat{H}^{JC}_{2}-\mathcal{E}_{0}(t)\mu_{01}(\vert
1;0\rangle\langle 0;0\vert+\vert 0;0\rangle\langle 1;0\vert )  \\ \nonumber
&&- \sum_{\ell=\pm }\mathcal{E}_{l,1}(t)\mu_{1,l}(
\vert 1;0\rangle\langle \ell;0\vert
+\vert \ell;0\rangle\langle 1;0\vert ),
\end{eqnarray}
where $\mathcal{E}_{0}(t)$ and $\mathcal{E}_{\pm,1}(t)$ are control
pulses to drive the system from the ground state $\vert
0;0\rangle$ to the state $\vert
1;0\rangle$ and the maximally entangled states $\vert\pm;0\rangle$, $\mu_{01}$ and $\mu_{1,\pm}=\pm\sqrt{2}/2\mu_{12}$ represent the transition dipole moments between four states. By applying the first-order Magnus expansion to the unitary evolution operator, the analytical wave function for the second polariton configuration via $2V$ excitation  can be obtained by  
\begin{eqnarray}\label{ctype2} \vert\psi_2^{(1)}(t)\rangle &=&\cos\theta_{0}(t)\vert0;0\rangle+\frac{i\theta_{0}^{*}(t)}{\vert\theta
_{0}\vert}\sin\theta_{0}(t)[\cos \theta _{1}(t)\vert 1;0\rangle\nonumber\\
&&+\frac{i\theta _{+,1}^{*}(t)}{\theta _{1}(t)}\sin \theta _{1}(t)\left\vert
+;0\right\rangle+\frac{i\theta_{-,1}^{*}(t)}{
\theta_{1}(t)}\sin\theta_{1}(t)\vert-;0\rangle],
\end{eqnarray}
where the complex pulse areas $\theta_{0}(t)=\mu_0\int_{t_{0}}^{t}dt^{\prime }\mathcal{E}_{0}(t^{\prime })e^{i\omega_{01}t^{\prime}}$ and $\theta_{\pm,1}(t)=\mu_{1,\pm}\int_{t_{0}}^{t}dt^{\prime }\mathcal{E}_{\pm,1}(t^{\prime })e^{i(\omega_{\pm,1}-\omega_{01})t^{\prime}}$, and $\theta_{1}(t)=\sqrt{\vert\theta_{+,1}(t)\vert^2+
\vert\theta_{-,1}(t)\vert^2}$.\\ \indent
By comparing Eq. (\ref{ctype2}) with the desired wave function of the target state in Eq. (\ref{WFp2}), the maximum degree of orientation in Eq. (\ref{cs2}) can be obtained with optimal amplitude conditions 
\begin{eqnarray}\label{CA2V}
\vert\theta_{0}(t_{f})\vert&=&\arccos\frac{\sqrt{10}}{6},\nonumber\\
\vert\theta_{\pm,1}(t_{f})\vert&=&\frac{1}{\sqrt{2}}\arccos\frac{3}{\sqrt{13}},
\end{eqnarray}
and  relative phase conditions
 \begin{eqnarray}\label{Cp2V}
 (\omega_{1,0}-\omega_{-,0})\arg[\theta_{0}(t_{f})]+\omega_{1,0}\arg[\theta_{-,1}(t_{f})]&=&\frac{\omega_{-,0}\pi}{2}+2g_{2}k\pi ,\nonumber\\
(\omega_{1,0}-\omega_{-,0})\arg[\theta_{+,1}(t_{f})]+(\omega_{+,0}-\omega_{1,0})\arg[\theta_{-,1}(t_{f})]&=&2g_{2}k\pi.
\end{eqnarray}\\ \indent
Using the method based on the designed pulse of the $V2$ excitation, we can obtain the relations of $\theta_0^{*}({t_f})=\mu_{01}A_{0}(\omega_{01})e^{i\phi_{10}(\omega_{01})}$ and $\theta_{\pm,1}^{*}(t_{f})=\mu_{1,\pm}A_{\pm}(\omega_{\pm,1})e^{i\phi_{\pm}(\omega_{\pm,0}-\omega_{01})}$ with the spectral amplitudes $A_{0}(\omega_{01})$ and $A_{\pm}(\omega_{\pm,1})$, and phases $\phi_{0}(\omega_{01})$ and $\phi_\pm(\omega_{\pm,1})$ at the transition frequencies $\omega_{01}$ and $\omega_{\pm,1}=\omega_{\pm,0}-\omega_{01}$. Here, we consider the three pulses used in the $2V$ excitation to have a Gaussian distribution, expressed as $A_{0}(\omega_{01})=A_{0}\text{exp}[{-(\omega-\omega_{01})^2/2(\Delta\omega)^2}]$ and $A_{\pm}(\omega_{\pm,0})=A_{\pm}\text{exp}[{-(\omega-\omega_{\pm,1})^2/2(\Delta\omega)^2}]$, where $A_{0}=\vert\theta_{0}(t_{f})\vert/\mu_{01}$ and $A_{\pm}=\vert\theta_{\pm,1}(t_{f})\vert/\mu_{1,\pm}$ represent the amplitudes.  The spectral phases are expanded as $\phi_{1,0}(\omega_{01})=\varphi_{1,0}-\omega\tau_{1,0}$ and $\phi_{\pm}(\omega_{\pm,1})=\varphi_{1,\pm}-\omega\tau_{1,\pm}$. By  performing  inverse Fourier transforms, the optimal control pulses  can be designed by
\begin{eqnarray} \label{w1e2}
\mathcal{E}_{0}(t)&=& \mathcal{E}_{0} e^{-\frac{(t-\tau_{1,0})^2}{2\tau^2}} \cos(\omega_{1,0}(t-\tau_{1,0}) + \varphi_{1,0}),\nonumber \\
\mathcal{E}_{\pm,1}(t)&=& \mathcal{E}_{\pm,1}e^{-\frac{(t-\tau_{1,\pm})^2}{2\tau^2}} \cos[(\omega_{\pm,0}-\omega_{1,0})(t-\tau_{1,\pm}) + \varphi_{1,\pm}],
\end{eqnarray}
with the strengths of the electric fields $\mathcal{E}_{0}=\sqrt{\frac{2}{\pi}} \frac{1}{\tau}\frac{\vert\theta_{0}\vert}{\mu_{01}}$ and $\mathcal{E}_{\pm,1}=\sqrt{\frac{2}{\pi}}\frac{1}{\tau}\frac{\vert\theta_{\pm,1}\vert}{\mu_{1,\pm}}$, the pulse duration is $\tau=1/\Delta\omega$. By controlling $\varphi_{1,\pm}-(\omega_{\pm,0}-\omega_{1,0})\tau_{1,\pm}=\phi_{\pm,0}-\frac{\pi}{2}$ and $\varphi_{1,0}-\omega_{1,0}\tau_{1,0}=\phi_{1,0}-\frac{\pi}{2}$, we can verify that the analytically designed control pulses in Eq. (\ref{w1e2}) satisfy both the amplitude and phase conditions in Eqs. (\ref{CA2V}) and (\ref{Cp2V}).
\section{RESULTS AND DISCUSSION  \label{RESULTS}}
We now examine the $V2$ and $2V$ control schemes in ultracold Carbonyl Sulfide (OCS) molecules, which were commonly used in experiments \cite{goban2008laser, trippel2014strongly,trippel2015two,karamatskos2019molecular}, with  $\mu=0.715$ D, $B=0.20286$ cm$^{-1}$, and the corresponding rotational period of $\tau_{0}=\pi/B=82.23$ps. We consider $g_{1}=g_{2}=0.1\omega_{01}=0.2B$ in the strong-coupling regime. To obtain the time-dependent wave functions $\vert\psi_1(t)\rangle$ and $\vert\psi_2(t)\rangle$ of two different polariton configurations described by the extended JC-model and driven by the analytically designed pulses, we numerically solve the corresponding Schr\"{o}dinger equation governed by the time-dependent Hamiltonian in the dipole approximation
\begin{eqnarray}\label{hp1n}
\hat{H}_{1}(t) &=&\hat{H}^{JC}_{1}- \sum_{\ell=\pm }\mathcal{E}_{1}(t)\tilde{\mu}_{0}(
\vert \ell;0\rangle\langle 0;0\vert+\vert 0;0\rangle\langle \ell;0\vert) \nonumber \\
&&-\mathcal{E}_{1}(t)\mu_{2,+}(\vert +;0\rangle\langle 2;0\vert +\vert 2;0\rangle \langle+;0\vert),
\end{eqnarray}
and
\begin{eqnarray} \label{hp2n}
\hat{H}_{2}(t) 
&=&\hat{H}^{JC}_{2}-\mathcal{E}_{2}(t)\mu_{01}(\vert
1;0\rangle\langle 0;0\vert+\vert 0;0\rangle\langle 1;0\vert )   \nonumber\\
&&- \sum_{\ell=\pm }\mathcal{E}_{2}(t)\mu_{1,l}(
\vert 1;0\rangle\langle \ell;0\vert
+\vert \ell;0\rangle\langle 1;0\vert ),
\end{eqnarray}
with the total control fields $\mathcal{E}_1(t)=\mathcal{E}_-(t)+\mathcal{E}_+(t)+\mathcal{E}_{2,+}(t)$ and $\mathcal{E}_2(t)=\mathcal{E}_0(t)+\mathcal{E}_{-,1}(t)+\mathcal{E}_{+,1}(t)$. Note that the interaction terms in Eqs. (\ref{hp1n}) and (\ref{hp2n}) are different from that in Eqs. (\ref{hp1}) and (\ref{hp2}), where we apply individual pulses to the corresponding transition paths separately. It implies that interactive effects between pulses, which are neglected in our analytical derivations, are considered in our numerical simulations.
\subsection{Maximizing the orientation for the first polariton configuration}
\begin{figure}\centering
\resizebox{0.7\textwidth}{!}{
\includegraphics{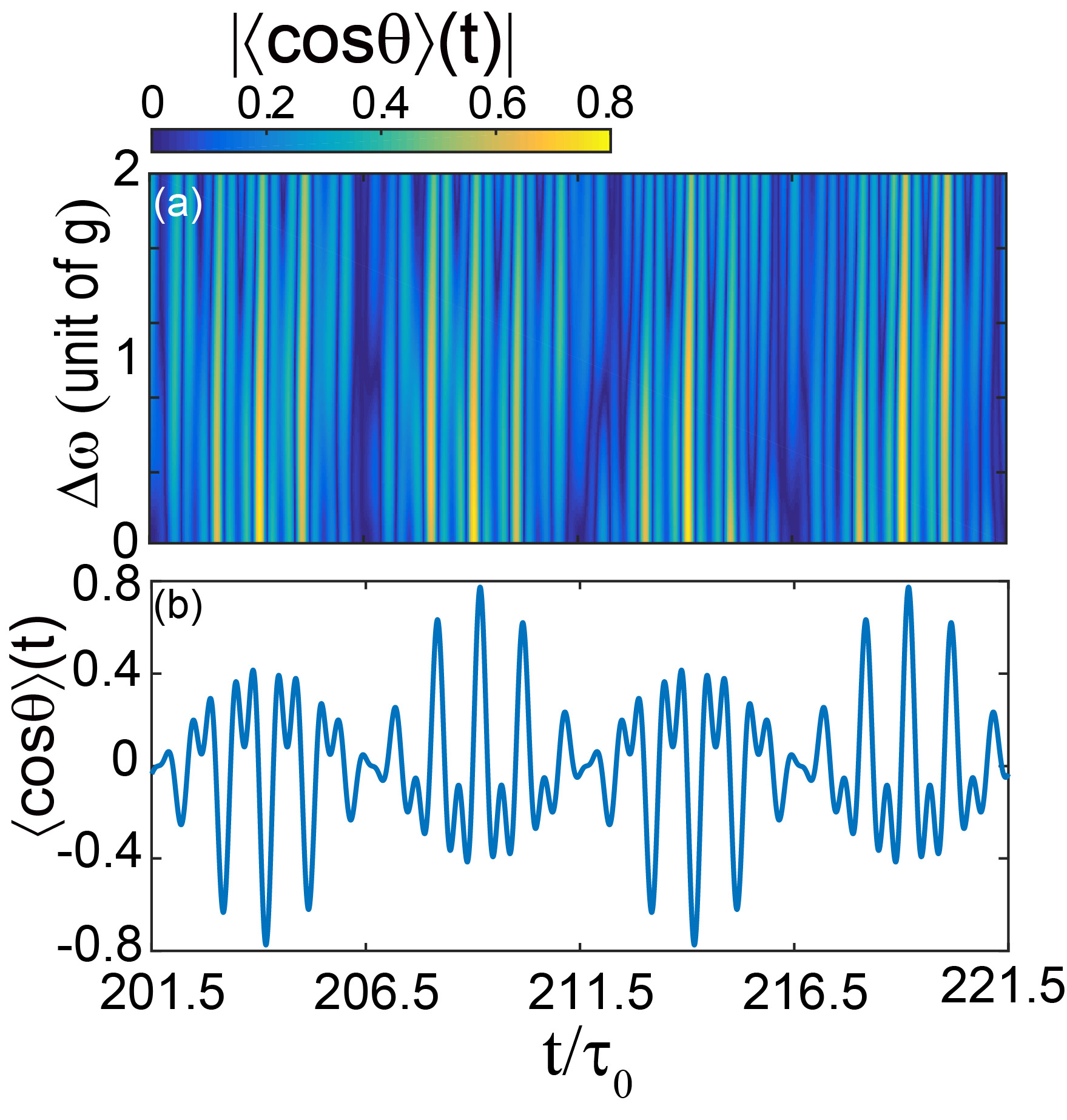}}\caption{The numerically calculated post-pulse orientation of the polariton driven by an optimized laser pulse. (a) The time-dependent degree of orientation $\vert\langle\cos\theta\rangle(t)\vert$ versus the bandwidth $\Delta\omega$ of the laser pulses. (b) $\langle\cos\theta\rangle(t)$ at a narrow bandwidth $\Delta\omega=0.1g_{1}$. }
\label{fig2}
\end{figure}
To examine the $V2$ control scheme for the first polariton configuration, we employ the analytically designed pulse as described in Eq. (\ref{w1e1}) to achieve the maximum orientation. 
For the numerical simulations, we select the amplitudes of the three pulses to satisfy the conditions in Eq. (\ref{am1}), and the phases $\varphi_{-,0}=\pi$, $\varphi_{+,0}=\pi/3$ and $\varphi_{2,+}=-\pi/3$ to satisfy the phase conditions outlined in Eq. (\ref{sp1}). 
Figure \ref{fig2}(a) demonstrates the time-dependent evolution of the  $\left|\langle\cos\theta\rangle(t)\right|$ relative to the bandwidth $\Delta\omega$ of  the analytically designed pulses with the center times $\tau_{\pm}=0$ and $\tau_{2,+}=121.053\tau_{0}$. The findings reveal that the pulse bandwidth significantly influences the molecular polariton orientation, and the orientation value can only reach the theoretical maximum value of $0.7746$ within a very narrow bandwidth range. As the bandwidth increases, the orientation's periodicity and maximum value are compromised. Figure \ref{fig2}(b) depicts the time-dependent  orientation $\langle\cos\theta\rangle(t)$ at $\Delta\omega=0.1g_{1}$. We notice that the orientation varies periodically with a revival period of $10\tau_{0}$, which agrees with the theoretical value. This period is significantly longer compared to considering only the lowest two rotational levels in the absence of the cavity, as discussed in \cite{fan2023quantum}.\\ \indent
\begin{figure}\centering
\resizebox{0.95\textwidth}{!}{
\includegraphics{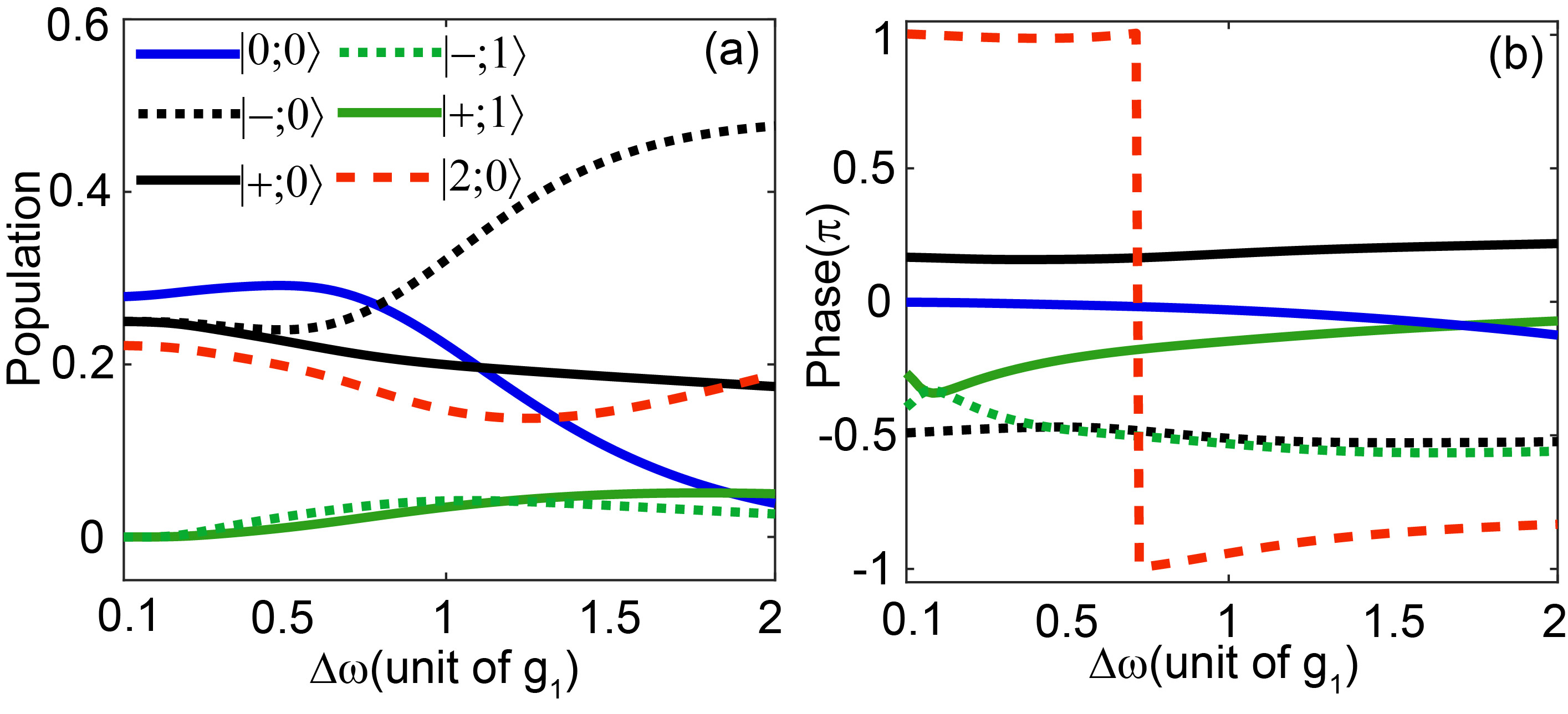} }\caption{The final populations (a) and phases (b) in states $\vert 0;0\rangle$, $\vert \pm;0\rangle$, $\vert \pm;1\rangle$, and $\vert 2;0\rangle$ versus the bandwidth $\Delta\omega$ of three laser pulses.
}
\label{fig3}
\end{figure}
To understand the impact of bandwidth $\Delta\omega$ on the maximum orientation value, Fig. \ref{fig3} plots the final populations and phases of the states $\vert 0;0\rangle$, $\vert \pm;0\rangle$, $\vert \pm;1\rangle$, and $\vert 2;0\rangle$ versus bandwidth $\Delta\omega$ of the analytically designed pulses, while keeping other parameters of the pulse the same as that in  Fig. \ref{fig2}. In the narrow bandwidth regime, the four states $\vert 0;0\rangle$, $\vert \pm;0\rangle$,  and $\vert 2;0\rangle$ are populated with optimal distributions of  $P_{0,0}=27.8\%$, $P_{\pm,0}=25\%$, and $P_{2,0}=22.2\%$. As the bandwidth increases, the populations in $\vert \pm;1\rangle$ in Fig. \ref{fig3}(a) become visible, and in state $\vert -;0\rangle$ gradually increases, whereas the population in states $\vert +;0\rangle$ and $\vert 2;0\rangle$ gradually decreases. By analyzing the transitions between molecular polariton states, we can see that the pulses with broad bandwidths can induce the transitions from $\vert \pm;0\rangle$ to $\vert \pm;1\rangle$  and from $\vert -;0\rangle$ to $\vert 2;0\rangle$. We can also observe from Fig. \ref{fig3}(b) that the phases of the states take place a significant shift compared to the optimal phase and no longer satisfy the optimal phase conditions in the broad bandwidth regime. \\ \indent
\begin{figure}\centering
\resizebox{0.85\textwidth}{!}{
\includegraphics{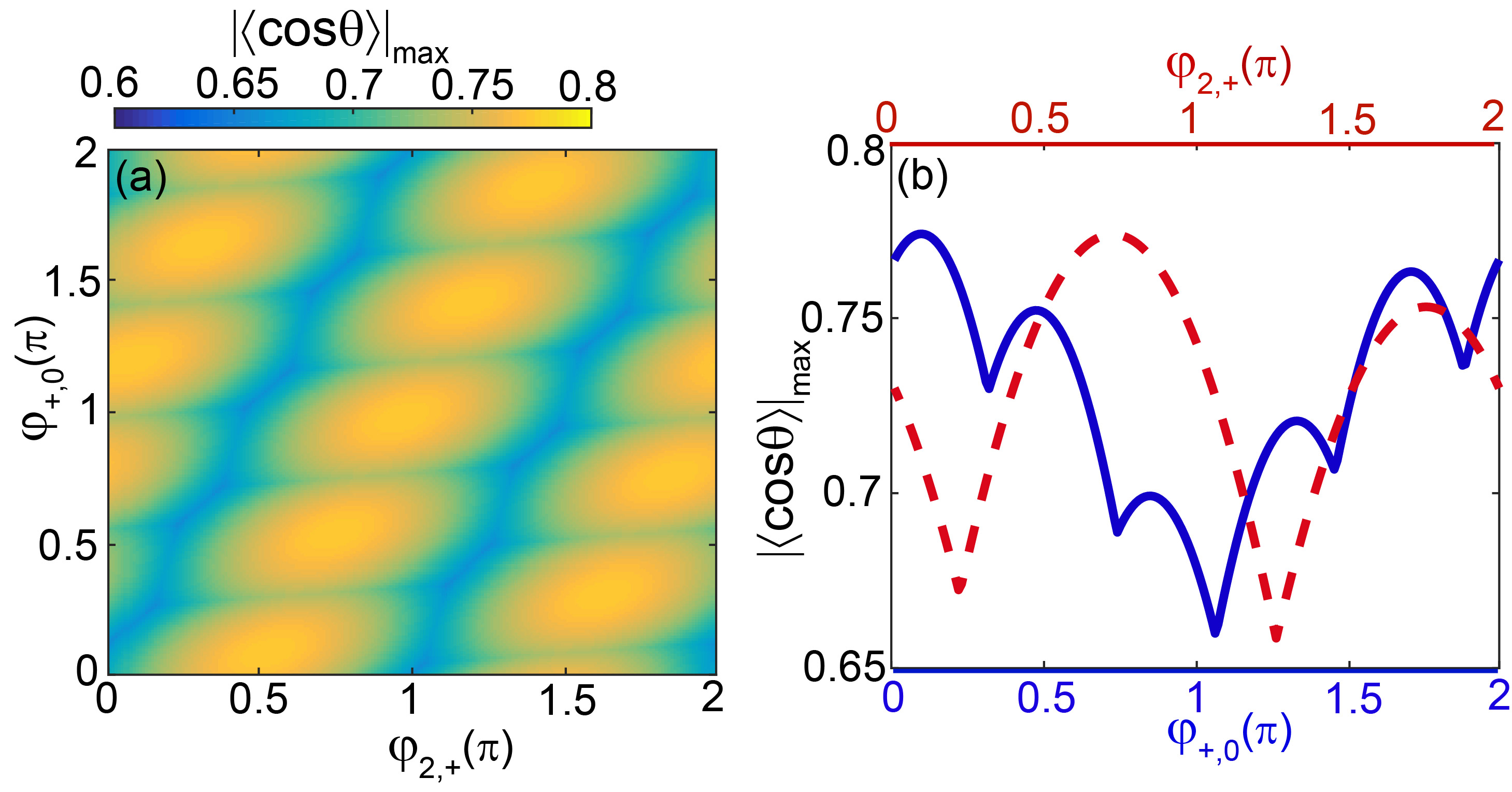} }\caption{(a) The maximum molecular polariton orientation versus the pulse phases $\varphi_{+,0}$ and $\varphi_{2,+}$  and (b) the corresponding cut lines at  $\varphi_{2,+} = 5\pi/9$ (blue solid line) and $\varphi_{+,0} = 5\pi/9$ (red dashed line), respectively. }
\label{fig4}
\end{figure}
To illustrate how the phases of the analytically designed pulses influence the maximum degree of orientation, Fig. \ref{fig4} plots the maximum orientation $\vert\langle\cos\theta\rangle\vert_{\text{max}}$ as a function of the two phases $\varphi_{+,0}$ and $\varphi_{2,+}$ with $\Delta\omega=0.1g_{1}$, while keeping other pulse parameters unchanged. We can see that the maximum orientation depends on the phases. The inconsistency to the phases $\varphi_{+,0}$ and $\varphi_{2,+}$  arises from the reason that the different transition frequency from the state $\vert 0;0\rangle$ to state $\vert +;0\rangle$ and from the state$\vert +;0\rangle$ to state $\vert 2;0\rangle$. To gain insights into the impact of the phases, Fig. \ref{fig4}(b) plots the cut lines of the maximum orientation at $\varphi_{2,+}=5\pi/9$  and $\varphi_{+,0}=5\pi/9$  as functions of $\varphi_{+,0}$ and $\varphi_{2,+}$, respectively. The  maximum orientation occurs  at $\varphi_{+,0}=\pi/9$ and  $\varphi_{2,+}=7\pi/9$, which aligns with the theoretical prediction by Eq. (\ref{sp1}).\\ \indent
\subsection{Maximizing the orientation for  the second polariton configuration}
\begin{figure}\centering
\resizebox{0.7\textwidth}{!}{
\includegraphics{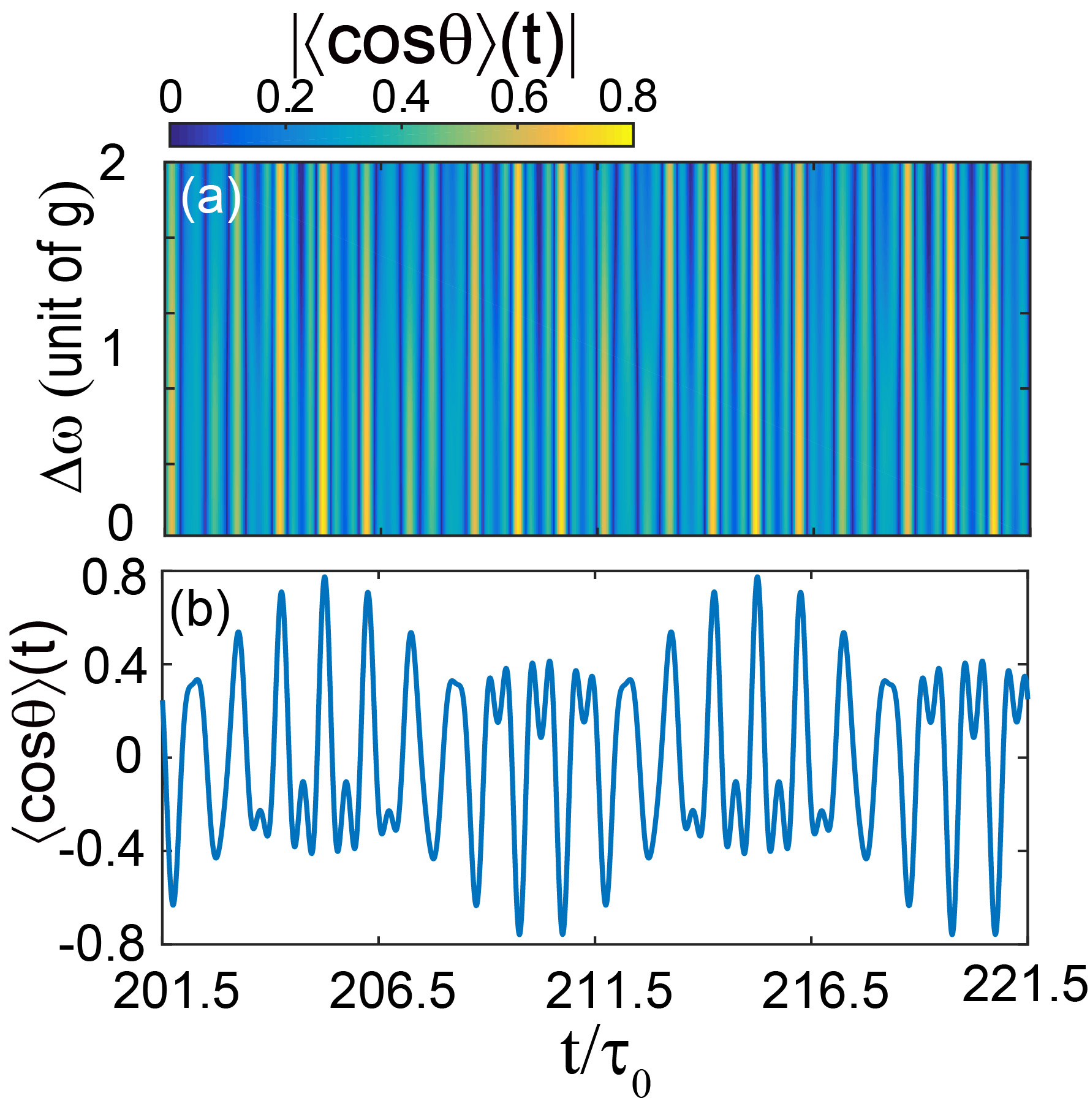} }\caption{The numerically calculated post-pulse orientation of the polariton driven by an optimized laser pulse. (a) The time-dependent degree of orientation $\vert\langle\cos\theta\rangle(t)\vert$ versus the bandwidth $\Delta\omega$ of the laser pulses. (b) $\langle\cos\theta\rangle(t)$ at a narrow bandwidth $\Delta\omega=0.1g_{2}$.}
\label{fig5}
\end{figure}
We now explore the second polariton configuration by utilizing the drive pulses described in Eq. (\ref{w1e2}) to examine the dependence of the maximum orientation on the bandwidth. In the simulation, three pulses are used with the delay times $\tau_{1,0}=0$ and $\tau_{1,\pm}=109.524\tau_{0}$. To meet the required amplitude and phase conditions, we apply the amplitude conditions outlined in Eq. (\ref{CA2V}) and fix the phase conditions to $\varphi_{1,0}=0$, $\varphi_{1,-}=31\pi/20+\phi_{\tau}$, and $\varphi_{1,+}=9\pi/20$, where $\phi_{\tau}=0.1905\pi$ is used to eliminate the phase shift caused by the delay time. Figure \ref{fig5} (a) shows the time-dependent evolution of the orientation $\vert\langle\cos\theta\rangle(t)\vert$ as a function of pulse bandwidth $\Delta\omega$. The orientation exhibits a periodic change over time. Unlike the first polariton configuration in Fig. \ref{fig2}(a), the orientation of the second polariton configuration remains close to $0.7746$ even as the bandwidth increases to $0.5g$, demonstrating high stability for bandwidth. In addition, Fig. \ref{fig5}(b) shows the time-dependent evolution of the orientation degree of the molecular polariton when the bandwidth is $\Delta\omega=0.1g_{2}$. Due to differences in energy level structure, this dynamic behavior differs from the pattern shown in Fig. \ref{fig2}(b). 
\begin{figure}\centering
\resizebox{0.95\textwidth}{!}{
\includegraphics{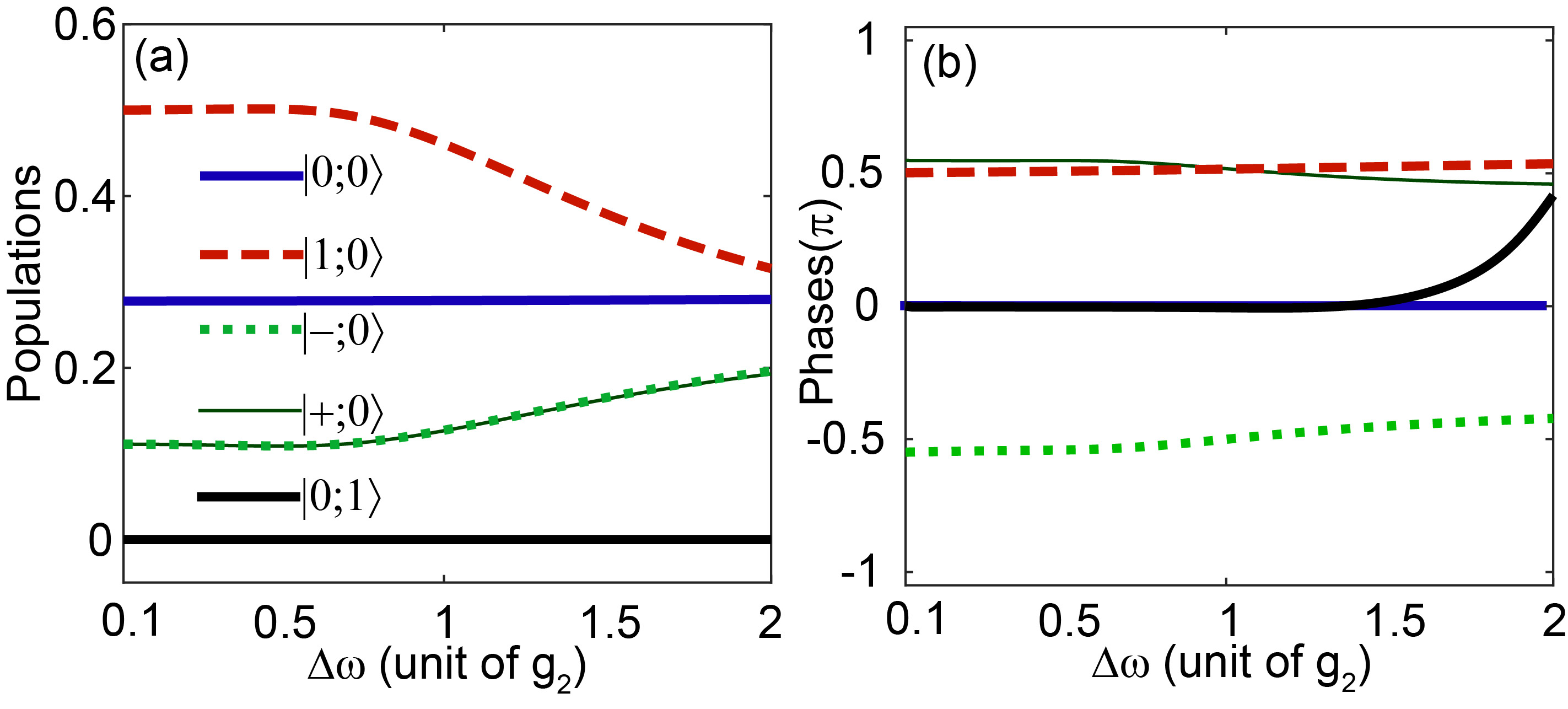} }\caption{The final populations (a) and phases (b) in states $\vert 0;0\rangle$ (blue solid line), $\vert -;0\rangle$ (green dashed line), $\vert +;0\rangle$ (green dashed line), $\vert 1;0\rangle$ (red dashed line), and $\vert 0;1\rangle$ (black solid line) versus the bandwidth $\Delta\omega$ of three laser pulses.
}
\label{fig6}
\end{figure}\\ \indent
To gain insight into the underlying excitation mechanism in the second polariton configuration, Fig. \ref{fig6} depicts the final populations and phases versus bandwidth for the states $\vert 0;0\rangle$, $\vert \pm;0\rangle$, $\vert 1;0\rangle$, and $\vert 0;1\rangle$. We find that the populations and phases exhibit different characteristics in different ranges of control field bandwidth. Specifically, for the bandwidth $\Delta\omega\in(0.1g, 0.5g)$, the four states $\vert 0;0\rangle$, $\vert \pm;0\rangle$, and $\vert 1;0\rangle$ achieve the optimal distributions of $P_{0,0}=27.8\%$, $P_{\pm,0}=11.1\%$, and $P_{1,0}=50\%$, respectively. As the bandwidth increases, the populations in $\vert\pm;0\rangle$ in Fig. \ref{fig6}(a) gradually increase, and in state $\vert 1;0\rangle$ gradually decreases, but the distribution of states remains within the four-state system, which differs from the first polariton configuration shown in Fig. \ref{fig3}(a). This result indicates the optical transitions via high-order terms in the Magnus expansion, which reduces the population distribution to deviate from the optimal distribution. Furthermore, as the bandwidth increases, the phase of the states also deviates slightly, no longer strictly maintaining the optimal phase, as shown in Fig. \ref{fig6}(b).\\ \indent
\begin{figure}\centering
\resizebox{0.95\textwidth}{!}{
\includegraphics{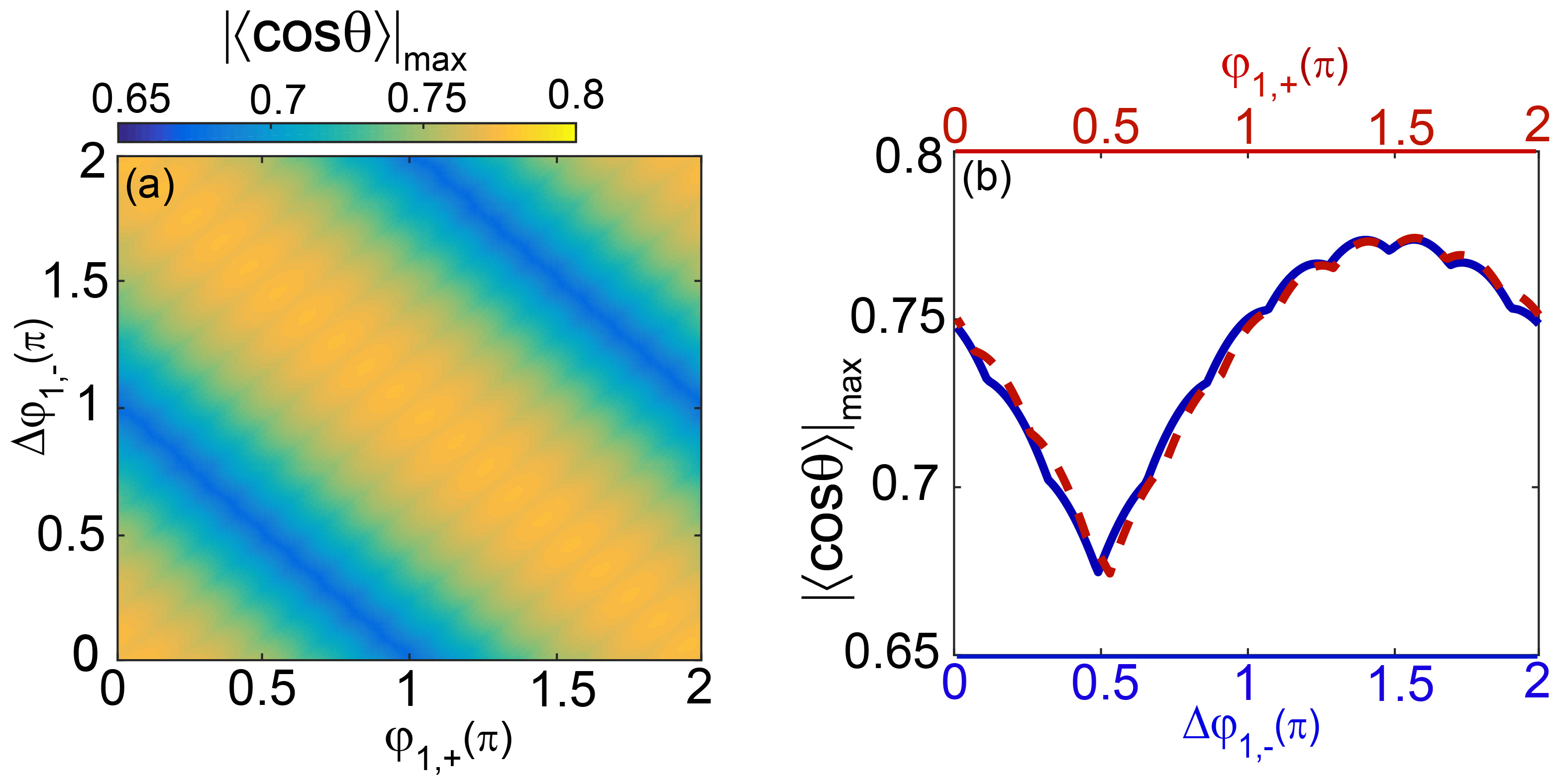} }\caption{(a) The maximum molecular polariton orientation versus the pulse phases $\varphi_{1,+}$ and $\Delta\varphi_{1,-}$  and (b) the corresponding cut lines at $\varphi_{1,+} = 0.55\pi$ (blue solid line) and $\Delta\varphi_{1,-} = 0.35\pi$ (red dashed line), respectively. }
\label{fig7}
\end{figure}
Figure \ref{fig7} shows how the maximum orientation depends on the phases $\varphi_{1,+}$ and $\Delta\varphi_{1, -}=\varphi_{1, -}-\phi_{\tau}$ for a pulse bandwidth $\Delta\omega=0.1g_{2}$, while keeping other pulse parameters consistent with Fig. \ref{fig5}. It can be observed that the periodicity in Fig. \ref{fig7} shows a significant difference compared to the periodicity observed in Fig. \ref{fig4}. This difference is primarily due to the different energy level structures of the two four-state systems. Notably, within the same phase interval, the second polariton configuration has more phase combinations to achieve the maximum orientation than the first. To give a clear view of the specific effect of phases on the maximum orientation, Fig. \ref{fig7}(b) presents the cut lines of the maximum orientation at $\varphi_{1,+}=0.55\pi$ and $\Delta\varphi_{1,-}=0.35\pi$ as functions of the phases $\Delta\varphi_{1,-}$ and $\varphi_{1,+}$, respectively. The results show that the maximum orientation occurs at $\Delta\varphi_{1,-}=1.45\pi$ and $\varphi_{1,+}=1.65\pi$, which is consistent with the theoretical prediction provided by Eq. (\ref{Cp2V}).
\section{conclution\label{conclution}}
We investigated the precise control of the maximum orientation for a single molecule within a cavity. To show the roles of the cavity-photon, we studied two distinct polariton configurations resulting from the strong coupling of the lowest three rotational states with a fundamental frequency cavity and a second harmonic cavity, respectively. To achieve the maximum degree of molecular orientation in the cavity, we proposed two different excitation schemes based on the two types of molecular polaritons. By utilizing the first-order Magnus expansion for the time-dependent unitary operator, we derived analytical time-dependent wave functions for the two polaritons. Subsequently, we obtained the two corresponding pulse-area theorems employed to design three optimal pulses for the two excitation schemes analytically. To examine these two pulse-area theorems, we applied these schemes to ultracold OCS molecules in their rotational ground state. Our numerical simulations demonstrated the validity of the two control schemes when using narrow bandwidth pulses. Finally, we assessed the sensitivity of the control schemes to the phases of the designed control pulses. Our results present a theoretical strategy for precisely controlling three-state molecular rotations within a cavity using different coupling approaches, offering valuable insights for designing coherent control experiments.
\section*{Acknowledgments}
This work was supported by the National Natural Science Foundation of China (NSFC)  under Grants 12274470 and 62273361, and the Hunan Province for Distinguished Young Scholars under Grant 2022JJ10070. This work was carried out in part using computing resources at the High Performance Computing Center of Central South University.
%\bibliography{reference}
%apsrev4-2.bst 2019-01-14 (MD) hand-edited version of apsrev4-1.bst
%Control: key (0)
%Control: author (8) initials jnrlst
%Control: editor formatted (1) identically to author
%Control: production of article title (0) allowed
%Control: page (0) single
%Control: year (1) truncated
%Control: production of eprint (0) enabled
%

\end{document}